\documentclass[a4paper,UKenglish,cleveref,autoref]{lipics-v2019}

\title{Tackling the Awkward Squad for Reactive Programming: The Actor-Reactor Model}
\titlerunning{Tackling the Awkward Squad for Reactive Programming}

\author{Sam Van den Vonder}{Software Languages Lab, Vrije Universiteit Brussel, Belgium}{}{https://orcid.org/0000-0002-9241-1098}{Research Foundation - Flanders (FWO) grant No. 1S95318N}

\author{Thierry Renaux}{Software Languages Lab, Vrije Universiteit Brussel, Belgium}{}{https://orcid.org/0000-0002-9301-2187}{Flanders Innovation \& Entrepreneurship (VLAIO) ``Cybersecurity Initiative Flanders'' program}

\author{Bjarno Oeyen}{Software Languages Lab, Vrije Universiteit Brussel, Belgium}{}{https://orcid.org/0000-0002-2100-4559}{Research Foundation - Flanders (FWO) grant No. 1S93820N}

\author{Joeri De Koster}{Software Languages Lab, Vrije Universiteit Brussel, Belgium}{}{https://orcid.org/0000-0002-2932-8208}{}

\author{Wolfgang De Meuter}{Software Languages Lab, Vrije Universiteit Brussel, Belgium}{}{https://orcid.org/0000-0002-5229-5627}{}

\authorrunning{S. Van den Vonder, T. Renaux, B. Oeyen, J. De Koster and W. De Meuter} 

 \Copyright{Sam Van den Vonder, Thierry Renaux, Bjarno Oeyen, Joeri De Koster and Wolfgang De Meuter}



\supplement{The code artefact of this paper can be found via DOI \url{https://doi.org/10.5281/zenodo.3862954} or via the corresponding artefact publication~\cite{darts-artifact}}

\nolinenumbers 

\EventEditors{Robert Hirschfeld and Tobias Pape}
\EventNoEds{2}
\EventLongTitle{34th European Conference on Object-Oriented Programming (ECOOP 2020)}
\EventShortTitle{ECOOP 2020}
\EventAcronym{ECOOP}
\EventYear{2020}
\EventDate{July 13--17, 2020}
\EventLocation{Berlin, Germany}
\EventLogo{}
\SeriesVolume{166}
\ArticleNo{19}

\usepackage{inconsolata}
\usepackage[T1]{fontenc}
\usepackage{hhline} 

\usepackage{tcolorbox}

\definecolor{line-numbers}{rgb}{0.3,0.3,0.3}
\definecolor{comment-color}{rgb}{0.4,0.4,0.45}
\definecolor{listing-bg-color}{rgb}{0.95,0.95,0.95}

\usepackage{listings}
\definecolor{lightgray}{rgb}{.9,.9,.9}
\definecolor{purple}{rgb}{0.65, 0.12, 0.82}

\definecolor{bluekeywords}{rgb}{0.0, 0.0, 0.7}
\definecolor{redsymbols}{rgb}{0.7, 0.0, 0.0}
\definecolor{darkgreenstrings}{rgb}{0.18,0.54,0.34}
\definecolor{darkgraycomments}{rgb}{0.5, 0.5, 0.5}

\lstdefinelanguage{Stella}{
  keywords={class, def-fields, def-constructor, set!, def, def-routine, def-method, new, actor, reactor, cond, else, if, def-stream, spawn-actor, out, tick, ror, spawn-reactor, react-to, emit, send, monitor},
  morecomment=[l]{//},
  morestring=[b]",
  morecomment=[s][\color{redsymbols}]{\ '}{\ },
  alsoletter={!, ?, \$, \%, &, *, -, +, ., /, :, <, =, >, @, ^, _, ~},
  keywordstyle=\color{bluekeywords},
  commentstyle=\color{darkgraycomments},
  stringstyle=\color{darkgreenstrings},
  sensitive=true,
  moredelim=[is][\color{redsymbols}]{|}{|}
}

\lstdefinelanguage{Scala}{
  morekeywords={abstract,case,catch,class,def,%
    do,else,extends,false,final,finally,%
    for,if,implicit,import,lazy,match,mixin,%
    new,null,object,override,package,%
    private,protected,requires,return,sealed,%
    super,this,trait,true,try,%
    type,val,var,while,with,yield},
  otherkeywords={=,=>,<-,<\%,<:,>:,\#,@},%
  sensitive=true,%
  morecomment=[l]//,%
  morecomment=[n]{/*}{*/},%
  morestring=[b]",%
  morestring=[b]',%
  morestring=[b]""",%
  keywordstyle=\color{bluekeywords},
  commentstyle=\color{darkgraycomments},
  stringstyle=\color{darkgreenstrings},
}%

\lstdefinelanguage{JavaScript}{
  keywords={break, case, catch, continue, debugger, default, delete, do, else, false, finally, for, function, if, in, instanceof, new, null, return, switch, this, throw, true, try, typeof, var, void, while, with, const},
  morecomment=[l]{//},
  morecomment=[s]{/*}{*/},
  morestring=[b]',
  morestring=[b]",
  ndkeywords={class, export, boolean, throw, implements, import, this},
  keywordstyle=\color{blue},
  ndkeywordstyle=\color{darkgray},
  identifierstyle=\color{black},
  commentstyle=\color{purple},
  stringstyle=\color{green},
  sensitive=true,
}

\newcolumntype{n}{>{\hsize=.35\hsize}X}
\newcolumntype{o}{>{\hsize=.15\hsize}X}

\begin{document}

\maketitle

\begin{abstract}
Reactive programming is a programming paradigm whereby programs are internally represented by a dependency graph, which is used to automatically (re)compute parts of a program whenever its input changes. 
In practice reactive programming can only be used for some parts of an application: a reactive program is usually embedded in an application that is still written in ordinary imperative languages such as JavaScript or Scala.
In this paper we investigate this embedding and we distill ``the awkward squad for reactive programming'' as 3 concerns that are essential for real-world software development, but that do not fit within reactive programming.
They are related to long lasting computations, side-effects, and the coordination between imperative and reactive code.
To solve these issues we design a new programming model called the Actor-Reactor Model in which programs are split up in a number of actors and reactors.
Actors and reactors enforce a strict separation of imperative and reactive code, and they can be composed via a number of composition operators that make use of data streams.
We demonstrate the model via our own implementation in a language called Stella.
\end{abstract}

\section{Introduction}
\label{sec:typesetting-summary}
\begin{figure}[t]
	\centering
  	\includegraphics[width=0.65\linewidth]{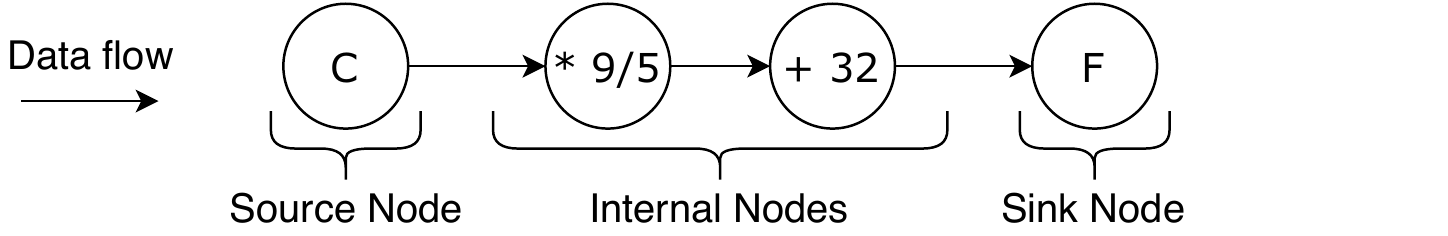}
  	\caption{Compiled structure of a reactive program to a DAG. Data flow is usually depicted from top to bottom, but depicted here from left to right.}
  	\label{fig:temperature_dag}
\end{figure}

Reactive programming is a programming paradigm that revolves around automatically changing the state of a program based on perpetually incoming values.
Historically, it was conceived as a solution to the problems of \emph{inversion of control} and \emph{callback hell} which occur when programming event-driven programs~\cite{DBLP:journals/csur/BainomugishaCCMM13}.
Reactive programming languages such as FrTime~\cite{DBLP:conf/esop/CooperK06}, Flapjax~\cite{DBLP:conf/oopsla/MeyerovichGBCGBK09}, Elm~\cite{DBLP:conf/pldi/CzaplickiC13} and REScala~\cite{DBLP:conf/aosd/SalvaneschiHM14} propose increasingly rich abstraction mechanisms to represent and compose so-called ``time-varying values'' or \emph{signals}, which are values that can change over time.
The ``Hello World!'' of reactive programming converts measurements of a Celsius thermometer to Fahrenheit. 
Given a signal~\texttt{C} that represents a changing temperature in Celsius, the expression \texttt{F = (C * 9/5) + 32} declares a new signal \texttt{F} that represents the temperature in Fahrenheit.
Changes to the value of \texttt{C} automatically give rise to a recomputation of \texttt{F}.
This is typically realised by compiling the reactive program into a directed acyclic graph (DAG), as exemplified in \cref{fig:temperature_dag}.

There are many incarnations of reactive languages and libraries built with various mainstream languages.
In academia, reactive languages include FrTime~\cite{DBLP:conf/esop/CooperK06} (Racket), Scala.React~\cite{Maier:176887} (Scala), REScala~\cite{DBLP:conf/aosd/SalvaneschiHM14} (Scala) and Flapjax~\cite{DBLP:conf/oopsla/MeyerovichGBCGBK09} (JavaScript). 
In industry, large companies such as Facebook develop and maintain frameworks such as ReactJS~\cite{reactjs} and React Native~\cite{reactnative} for reactive Graphical User Interfaces (GUIs), ReactiveX is a specification for reactive streams implemented in over 18 languages~\cite{reactivex} some of which are developed or maintained by companies such as Microsoft and Netflix, and an implementation of the ``Reactive Streams'' specification has been included in Java since version~9~\cite{reactivestreams, javaspec:jep266}.
Results from an empirical study on program comprehension suggest favourable results for reactive programming when compared to the Observer pattern in object-oriented programming~\cite{DBLP:journals/tse/SalvaneschiPANM17}.

In the rest of the paper we will abstract over the details of particular reactive programming languages and talk about them in terms of their role in the canonical DAG model for reactive programming.
\emph{Source nodes} correspond to the ``input'' signals of the reactive program.
Their values are typically provided by code that is external to the reactive program.
\emph{Internal nodes} are composed signals that constitute the reactive program.
\emph{Sink nodes} correspond to the ``output'' signals of the reactive program that constitute the program output.

Reactive programming languages and frameworks focus on the design and concepts of the \emph{internal part} of a reactive program.
Explained in terms of the DAG, they focus on language features and abstractions whereby programmers can express DAGs as easily and as declaratively as possible. 
Reactive programs also have 2 other parts: an \emph{input part} provides input to the reactive program by modifying its source nodes, and another (possibly different) \emph{output part} processes the output of the reactive program.
For example, in the temperature converter of \cref{fig:temperature_dag}, the source \texttt{C} may be connected to an input field in the GUI, and the result \texttt{F} may be displayed in the same GUI.
In another application, \texttt{C} may be connected to a distributed web-based data stream whereby the input of the reactive program is produced by an entirely different machine (e.g.~a weather station).

One cannot help but observe that reactive programming is only used to implement the internal part of a reactive program, and that it is usually embedded in an application that is still written in ordinary imperative languages such as JavaScript or Scala.
The main problem tackled in this paper is that the parts of existing reactive programming languages and frameworks that are responsible for input and output are ill-defined: They use ad-hoc mechanisms that can violate the invariants of reactive programs.
This is especially problematic for long lasting computations that block the reactive program~\cite{boner2014reactive}, and computations with side-effects (e.g.~modifying or rendering a GUI)~\cite{DBLP:conf/oopsla/Edwards09, Maier:176887}.
Mechanisms to embed reactive code between the imperative input and output parts of applications are poorly investigated in literature.
To this end, we have 2 main contributions.

\begin{enumerate}
	\item In analogy with the ``awkward squad'' for functional programming which are a set of application concerns that are essential for real-world software development, but that do not fit within the purely functional programming paradigm~\cite{Jones01tacklingthe}, in \cref{sec:problem-statement} we identify the \emph{``awkward squad for reactive programming''}.
		They are (1) long lasting computations, (2) embedding imperative code in reactive code, and (3) embedding reactive code in imperative code.
	\item Just like functional programming solves the problem by evacuating the awkward squad to a different location (i.e. monadic) of the program, we propose a programming model that solves these issues.
		First, in \cref{sec:object-oriented-subset} we define a class-based object-oriented data model that will guarantee that reactive programs cannot execute imperative code and long lasting computations.
		Second, in \cref{sec:actor-reactor-model} we introduce the \emph{Actor-Reactor Model} whereby the imperative input \& output parts of reactive programs must be modelled as \emph{actors}, and the internal parts of reactive programs are modelled as \emph{reactors}.
		Together, actors and reactors enforce that imperative and reactive code remains separated, but can still co-exist within the same application.
		To clearly demonstrate the concepts of this model we have designed and implemented an experimental language called Stella. 
\end{enumerate}

\section{Identifying the Awkward Squad for Reactive Programming}
\label{sec:problem-statement}

In this section we analyse the problems that may occur when embedding a reactive programming model in an imperative programming language, or when allowing the internal nodes of a reactive program to be programmed with the full power of a Turing-complete language.

\subsection{Long Lasting Computations}
\label{subsec:long-lasting-computations}
\label{program-termination-checking}
\begin{lstlisting}[language=Scala, caption={REScala reactive program that matches a regular expression with an input string.},label=lst:regex-long-lasting-computation, float]
val userInputSignal = Var("") // initial signal value is ""
val matches = Signal { "(A+)*B".r.findFirstMatchIn(userInputSignal()) }
\end{lstlisting}

Responsiveness is 1 of the 4 key properties of reactive systems outlined in the so-called Reactive Manifesto~\cite{boner2014reactive}\cite[Chapter~1]{Kuhn:2017:RDP:3153584}: ``\textit{Responsive systems focus on providing rapid and consistent response times, establishing reliable upper bounds so they deliver a consistent quality of service.}''~\cite{boner2014reactive}
However, reactive languages and frameworks often impose little to no restrictions on the types of expressions that can be used within a reactive program.
It is often easy to write code that accidentally makes the program no longer reactive.
For example, consider the reactive program in \cref{lst:regex-long-lasting-computation} (written in REScala) that checks whether user-provided input strings match a regular expression.
Line~1 defines a new source node called \texttt{userInputSignal}, and Line~2 defines an internal node called \texttt{matches} that is derived from the source node. 
Whenever the value of \texttt{userInputSignal} changes, the value of \texttt{matches} reflects whether the new string matches the regular expression \texttt{(A+)*B} (e.g.~\texttt{AB} and \texttt{AAAB}, but not \texttt{AAA}).
This program has a worst-case complexity of \texttt{$\mathcal{O}(2^n)$} with \texttt{n} the size of the input string.
Matching the string \texttt{AAAAA} fails after 112 steps, and matching 50 \texttt{A}'s fails only after \textasciitilde3 quadrillion steps~\cite{regexegg}.
This program clearly cannot be called reactive in the aforementioned sense.
While this specific example may be a contrived case of catastrophic backtracking in regular expressions, a developer can easily and accidentally introduce computations into reactive programs that (occasionally) have unintended consequences for their execution time.
We call this the \textbf{Reactive Thread Hijacking Problem}, because long lasting computations can ``hijack'' the thread of execution of a reactive program, thereby stopping the reactive program from being able to react to new input.

\begin{tcolorbox}[boxrule=1pt,arc=0pt,outer arc=0pt]
The language design problem that needs to be solved is how to ensure that a ``reliable upper bound'' can be imposed on long lasting computations within reactive programs.
\end{tcolorbox}

The Reactive Manifesto is intentionally vague about how to achieve this.
We identified 3 levels of reactivity that provide different program termination guarantees.

\subsubsection{Weak Reactivity}
The set of programs that we call \emph{weakly reactive} are those for which there are no guarantees with respect to how long they will take to execute.
This is usually because reactive programs can be programmed with the full power of a Turing-complete language.
Programs written in reactive languages such as FrTime~\cite{DBLP:conf/esop/CooperK06}, Flapjax~\cite{DBLP:conf/oopsla/MeyerovichGBCGBK09}, Elm~\cite{DBLP:conf/pldi/CzaplickiC13}, REScala~\cite{DBLP:conf/aosd/SalvaneschiHM14} and AmbientTalk/R~\cite{DBLP:conf/tools/CarretonMCM10}, or frameworks such as ReactJS~\cite{reactjs}, ReactiveX~\cite{reactivex} and Akka Streams~\cite[Chapter~13]{roestenburg2015akka} all fall into this category.

\subsubsection{Eventual Reactivity}
The set of programs that we call \emph{eventually reactive} are those for which it can be proven that the program will eventually terminate.
There are both static and dynamic approaches to enforce program termination, each with varying levels of restrictions that they impose on the underlying program.
Static enforcement includes work such as total functional programming~\cite{DBLP:journals/jucs/ATurner04}, primitive recursion~\cite[Chapter~3]{brainerd1974theory}, and model checkers such as TERMINATOR~\cite{DBLP:conf/pldi/CookPR06}.
Techniques that dynamically enforce program termination can make use of both static information and run-time values, such as size-change termination~\cite{DBLP:conf/pldi/NguyenGTH19}.

\subsubsection{Strong Reactivity}
The set of programs that we call \emph{strongly reactive} are those for which the execution time of the reactive program does not depend on the size of its input.
In other words, the asymptotic worst-case complexity of the reactive program is guaranteed to be in $\mathcal{O}(1)$.
Reactive programs that are strongly reactive will never be unexpectedly slow because of certain types of input.
We consider this to be the strongest form of reactivity, but the trade-off is that the types of programs that can be written is severely restricted.
		
Synchronous programming languages such as Esterel~\cite{DBLP:journals/scp/BerryG92}, Lustre~\cite{DBLP:journals/tse/HalbwachsLR92}, C\'eu~\cite{DBLP:journals/tecs/SantAnnaIRRB17} and Signal~\cite{DBLP:conf/fpca/GautierG87} apply reactive programming to real-time systems~\cite{DBLP:conf/ifip/Berry89}. 
They rely on the assumption of the \textit{synchronous hypothesis}, where the output of the program is conceptually synchronous with its input and ``instantaneous''~\cite{DBLP:reference/crc/SimoneTP05}.
In contrast with strong reactivity this is not a constraint on program complexity or execution time, but on the correctness of event processing whereby 1 event must be completely processed by the reactive program before the next event is considered.
Constant-time (non-reactive) programming languages such as FaCT~\cite{DBLP:conf/secdev/CauligiSBJHJS17} and Jasmin~\cite{DBLP:conf/ccs/AlmeidaBBBGLOPS17} are concerned with bounding execution times of certain instruction to prevent leaking secret information (encryption keys) based on execution time.
Constant-time programming is not a restriction of program complexity, but a restriction on the execution time of specific instructions, which may not vary in function of sensitive run-time information.
ActiveSheets~\cite{DBLP:conf/ecoop/VaziriTRSH14} is a reactive language where programs consist of Microsoft Excel spreadsheets.
While many spreadsheet operations are strongly reactive (e.g.~arithmetic), there are exceptions such as \texttt{SEARCH} and \texttt{REPLACE} for searching and replacing substrings. 
Finally, the RT-FRP language~\cite{DBLP:conf/icfp/WanTH01} is a statically typed reactive programming language where the time and space (in memory) cost for each execution step is statically bound.
To the best of our knowledge this language is strongly reactive.

\subsubsection{Summary}

\begin{figure}[t]
	\centering
  	\includegraphics[width=0.65\linewidth]{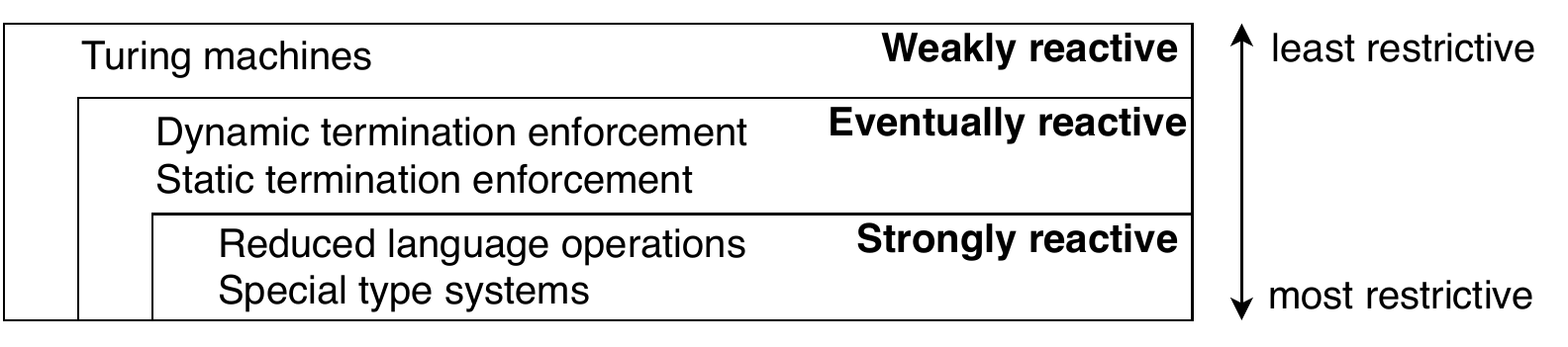}
  	\caption{Termination guarantees for different levels of reactivity.}
  	\label{fig:long_lasting_computations_expressivity}
\end{figure}

Expressions in existing reactive languages and frameworks can accidentally or unintentionally cause long lasting computations that block the reactive program, thus turning the program no longer reactive.
We identified 3 different levels of reactivity that are summarised in \cref{fig:long_lasting_computations_expressivity}, together with the possible techniques on how to enforce them.
Stricter enforcement of program termination will result in reactive programs that have stronger guarantees with respect to the processing of their input, i.e. a more reliable upper bound can be placed on their execution time.
However, the trade-off is that the types of programs that can be written is also reduced with stricter enforcement.
It is up to developers to choose a reactive language or framework that can enforce a particular level of reactivity that is appropriate for the application at hand, i.e. eventually reactive or strongly reactive.

\subsection{Embedding Imperative Code in Reactive Code}
\label{subsec:effectful-computations}

\begin{lstlisting}[language=Scala, caption={REScala reactive program with side-effects.},label=lst:rescala-side-effects, float]
val counter = Var(0) // initial source value is 0
Signal { print("A" + counter() + " ") }
Signal { print("B" + counter() + " ") }
\end{lstlisting}

Effectful computations are extremely tricky to understand and debug when they are embedded within the nodes of a dependency graph~\cite{DBLP:conf/oopsla/Edwards09, Maier:176887}.
Consider the reactive program (written in REScala) in \cref{lst:rescala-side-effects}. 
Line~1 defines a new source node called \texttt{counter}, and Lines~2 and~3 define two internal nodes that print either \texttt{A} or \texttt{B} to the console, followed by the value of the counter.
When this program is executed, the initial value of \texttt{counter} is propagated through the program, and \texttt{``A0 B0''} is printed to the console in the order of evaluation (from top to bottom).
However, when the value of \texttt{counter} is updated to 1, approximately 50\% of the time the program output is reversed and \texttt{``B1 A1''} is printed. 
We call this problem the \textbf{Reactive Update Order Leak}, because effectful computations leak information about the update order of subexpressions within reactive programs, and their correct execution may even rely on a specific order.

The update order of the DAG is usually not part of the semantics of a reactive program.
Reactive programming languages such as FrTime~\cite{DBLP:conf/esop/CooperK06}, Flapjax~\cite{DBLP:conf/oopsla/MeyerovichGBCGBK09} and REScala~\cite{DBLP:conf/aosd/SalvaneschiHM14} prevent \textit{glitches} (temporary inconsistencies in the program~\cite{DBLP:conf/esop/CooperK06}) by specifying that updates should be executed in a topological order of the dependency graph.
Some implementations parallelise the execution of certain regions of the DAG, such as the conceptual propagation model of Elm~\cite{DBLP:conf/pldi/CzaplickiC13} and a parallel version of the REScala update algorithm~\cite{DBLP:journals/pacmpl/DrechslerMSM18}.
Streaming frameworks such as ReactiveX~\cite{reactivex} and Akka Streams~\cite[Chapter~13]{roestenburg2015akka} do not feature such an algorithm, and instead only specify that parent nodes should be updated before their child nodes.

All of the aforementioned technologies allow multiple valid update orders to be used for a given program.
This is good for language implementers, because it gives them a lot of freedom to tweak and optimise (e.g.~parallelise) how values propagate through the reactive program.
However, for application developers this means that the concrete update order can vary across different implementations or versions of the same language or framework.
The order can even change at run-time, which is the case in our experiments with the code snippet of \cref{lst:rescala-side-effects}, where the execution of the program yields nondeterministic results.

The root of the problem are unconstrained side-effects in interactive applications.
Not only can side-effects cause bugs that are difficult to find and reproduce because of an unlucky ordering in some propagations through the DAG, but they are also very difficult to coordinate and have a detrimental effect on behavioural composition~\cite{coherentreaction}. 
Recognising these issues, most reactive programming languages and frameworks already forbid side-effects within the DAG of a reactive program, either through language design or via programmer guidelines (e.g.~REScala guidelines~\cite[1.5.3]{rescala-side-effects}).
However, as we will discuss in \cref{subsec:coordination-of-paradigms}, they are rarely successful in banishing side-effects completely.

\begin{tcolorbox}[boxrule=1pt,arc=0pt,outer arc=0pt]
The language design problem that needs to be be solved is how to allow the coexistence of effectful computations that react to values arriving at a certain internal node of the dependency graph without accidentally or nondeterministically affecting the behaviour of effectful computations that reside in other nodes of the dependency graph.
\end{tcolorbox}

\subsection{Embedding Reactive Code in Imperative Code}
\label{subsec:coordination-of-paradigms}
Programs written in existing reactive programming languages and frameworks are not subject to the Reactive Update Order Leak (\cref{subsec:effectful-computations}) when their code is \emph{purely} functional.
However, we observe that this requirement is not met in practice, because parts of real-world programs often depend on long lasting computations and effectful computations.
For instance, input can come from a GUI or be streamed in from another computer over a network connection, and output may be used to modify a GUI or to push a notification to a user.
A complete solution must therefore be able to bridge what we call the \textbf{Reactive/Imperative Impedance Mismatch}, in analogy with the Object-Relational Impedance Mismatch for object-oriented programming.

Because the embedding of reactive code within imperative code is unavoidable in real-world reactive applications, existing reactive languages and frameworks all tackle the Reactive/Imperative Impedance Mismatch to some degree. 
We argue that their solutions either have limited applicability, or are ad hoc solutions with unclear semantics. 
What follows is a non-exhaustive list of mechanisms that we could identify in related work. 

\begin{description}
	\item[Domain specific features] are special language features tailored to a specific domain, usually involving GUIs.
		For instance, Flapjax~\cite{DBLP:conf/oopsla/MeyerovichGBCGBK09}, Elm~\cite{DBLP:conf/pldi/CzaplickiC13}, and ReactJS~\cite{reactjs} provide a DSL for building GUIs whereby source nodes are automatically created and updated by GUI components, and the GUI automatically integrates with sink nodes.
    Such languages typically also feature built-in domain specific signals with narrow applicability, such as Elm's \texttt{Mouse.position} signal~\cite{DBLP:conf/pldi/CzaplickiC13}.
	
	\item[Special forms] are built-in language constructs with special evaluation rules.
    Reactive languages often offer special forms to provide operators that cannot be implemented from within the language itself.
		For example, Elm~\cite{DBLP:conf/pldi/CzaplickiC13} offers a built-in \texttt{syncGet} operation to execute synchronous HTTP requests.
		To prevent this computation from blocking the reactive program, Elm also offers an \texttt{async} special form to execute operations in the background.
	
	\item[Metaprogramming] involves mechanisms to manually construct or modify parts of the reactive program.
		For example, FrTime and REScala offer built-in primitives to create and (destructively) modify source nodes of the reactive program.
		The semantics of manual assignments to source nodes are often unclear, especially when multiple threads of execution are involved. 
		Streaming frameworks such as ReactiveX~\cite{reactivex} and Akka Streams~\cite[Chapter~13]{roestenburg2015akka} offer built-in operators to define
    new streams ex nihilo, reducing their dependence on special forms.
    Still, if there is no built-in support for a specific data source, a programmer has to open up the implementation of operators to carefully craft a new one.
	
	\item[Hidden concurrency] involves mechanisms whereby multiple threads of execution are used, but where the concurrent nature of the operations is hidden from the programmer.
		For example, in FrTime a dedicated thread manages the execution of the reactive program behind the scenes, while the Racket Read-Eval-Print Loop continues running on the main thread.
    From the REPL, a Scheme program can ``send'' input values to the reactive thread.
		Conversely, the output of FrTime programs can be reactively displayed in the output of the REPL.
    However, multi-threading and concurrency-control are not part of the FrTime programming model.
	
	\item[Periodic polling] enables embedding by periodically updating the source values of a reactive program from some computational process that is external to the reactive program.
    The quintessential example of this approach is the \texttt{seconds} behaviour as found in FrTime~\cite{DBLP:conf/esop/CooperK06}.
		The \texttt{seconds} behaviour is a free variable in the reactive program that is automagically updated to the current Unix time.
    Typically, dependents of the \texttt{seconds} behaviour use the timestamp only to determine whether they should imperatively poll an application-specific data source.
    Such code lets the reactive runtime schedule imperative code, giving rise to Reactive Update Order Leaks.

\end{description}

\begin{tcolorbox}[boxrule=1pt,arc=0pt,outer arc=0pt]
	The language design problem that needs to be solved is how to design a reactive programming language that supports the features of the awkward squad, but without introducing the Reactive Thread Hijacking Problem and the Reactive Update Order Leak. 
\end{tcolorbox}

In other words, it is important that imperative code and reactive code can coexist within the same application in such a way that the imperative code cannot accidentally violate the invariants of the reactive code.
The mechanisms employed by existing languages and frameworks are often highly specialised, have unclear semantics with regards to their interactions with the reactive program, and they do not solve these problems.

\subsection{Solution: General Idea}
\label{subsec:solution-general-idea}
The idea is to embrace that reactive applications always consist of both imperative and reactive parts.
Both are desired, and they complement each other~\cite{DBLP:conf/oopsla/Edwards09}. 
We give each their own thread and effect-set, and design simple composition operators to link them together.

The result is called the Actor-Reactor Model, which can be summarised as follows.
\begin{itemize}
	\item \emph{Actors} are used to represent the imperative input and output parts of reactive programs.
		They imperatively manage their own state and input-output, and are solely responsible for executing long lasting computations and effectful computations.
	\item \emph{Reactors} are used to encapsulate reactive programs, each with their own DAG and update thread. 
		By construction it is impossible for reactors to perform long lasting computations and side-effects.
	\item Coordination of actors and reactors happens via the \emph{data streams} which they consume and produce.
		We define a set of composition operators whereby the streams defined by actors can be linked to the source nodes of reactors and vice-versa how actors can act on changes to the sink nodes of reactors.
\end{itemize}

We have implemented the Actor-Reactor Model in an experimental language called Stella, which can be roughly divided into two levels.
First, an object-oriented base language is used to restrict reactors from producing side-effects and performing long lasting computations (\cref{sec:object-oriented-subset}). 
Second, actors and reactors are used to strictly separate the imperative and reactive parts of programs (\cref{sec:actor-reactor-model}). 
We have written a prototypical Continuation-Passing Style interpreter for Stella in TypeScript (in the style of~\cite[Chapter 5]{DBLP:books/daglib/0020601}).

\section{Base Language: OOP with Effect and Termination Guarantees}
\label{sec:object-oriented-subset}
In this section we will focus on the object-oriented base language of Stella.
Throughout all code examples we will consistently syntax highlight special forms (expressions with special evaluation rules) in blue. 
Strings are surrounded by double quotes and are highlighted in green. 
Symbols start with a single quote and are highlighted in red.

Stella is a dynamically typed language where all run-time values are objects.
Its native objects are booleans, numbers, strings, symbols and the value \texttt{\#undefined}, which are instances of the classes \texttt{Boolean}, \texttt{Number}, \texttt{String}, \texttt{Symbol} and \texttt{Undefined} respectively.
A program in Stella consists of 3 sets of top-level definitions: a set of classes, a set of \textit{actor behaviours} (``the class of an actor''), and a set of \textit{reactor behaviours} (``the class of a reactor'').

\subsection{Basic Expressions and ``Hello World!''}
Each Stella program must contain an actor behaviour called \texttt{Main}.
This actor behaviour must have a constructor named \texttt{start}.
To start a Stella program, the Stella runtime spawns an instance of the \texttt{Main} actor behaviour and invokes the \texttt{start} constructor.
Consider the \emph{``Hello World!''} program in \cref{lst:main-program} that defines such a behaviour. 
The body of its \texttt{start} constructor contains a single \texttt{println} statement in operator prefix notation (Polish Notation). 
This can be seen as synchronously sending a \texttt{println} message with no arguments to the \texttt{``Hello World!''} object of class \texttt{String}. 
Similarly, \texttt{(+ 1 2)} can be seen as sending a~\texttt{+} message to the object representation of the number~1 with one argument which is the object corresponding to number~2.
In the rest of this paper we will use the terminology of ``calling'' or ``invoking'' a method rather than sending synchronous messages (cf.\ SmallTalk~\cite{DBLP:books/aw/GoldbergR83}).

\begin{lstlisting}[language=Stella, caption={A ``Hello World!'' program in Stella.},label=lst:main-program, float]
(actor Main
  (def-constructor (start)
    (println "Hello World!")))
\end{lstlisting}

Expressions that may be used in the body of constructors and methods follow S-expression syntax~\cite[Chapter~1]{DBLP:books/mit/AbelsonS96} and are shown in \cref{lst:basic-expressions}. 
Local variables are introduced via the \texttt{def} special form (similar to \texttt{define} in Scheme), assignments use the \texttt{set!} special form, conditionals use the \texttt{if} and \texttt{cond} special forms. 
Two methods to test for equality are implemented by the superclass of all classes (\texttt{Object}): \texttt{equal?} tests for object equality, \texttt{eq?} for object reference equality.
All values are \texttt{\#true} except for \texttt{\#false} and \texttt{\#undefined}.

\begin{lstlisting}[language=Stella, caption={Examples of basic expressions in Stella},label=lst:basic-expressions, float]
(def hello "Hey")
(set! hello "... from the other side")
(if (equal? hello "Hey") (println "yes") (println "no")) // console prints "no"
\end{lstlisting}

\subsection{Abstract Data Types}
\label{subsec:abstract-data-types-size-change-termination}
Code in Stella can be either imperative or reactive. Both kinds of code can operate on the same data, but the allowed sets of operations differ.
Whereas imperative code may use the full power of a Turing-complete language, reactive code is restricted to operations that are guaranteed to terminate and that are free from side-effects.
To this end, abstract data types in Stella are represented by classes which may define local fields, constructors, methods, as well as  \emph{routines}.
Routines are special kinds of methods whose expressive power is a strict subset of that of methods. 
Routines have the following properties.

\begin{enumerate}
	\item Routines have no side-effects.
	\item Routines always terminate.
	\item Routines can only invoke other routines.
\end{enumerate}

To explain how we can enforce those properties, consider the \texttt{Pair} class defined in \cref{lst:classes} which can be used to represent linked lists.
Local fields of the class are declared on Line~2.
Line~4 defines a constructor called \texttt{initialize-with} with 2 formal parameters called \texttt{initial-car} and~\texttt{initial-cdr} that will initialize the fields of a new pair.
Lines~8--9 define two routines called \texttt{first} and \texttt{second} with no arguments which are ``getters'' for the \texttt{car} and \texttt{cdr} field, and correspondingly, Lines~10 and~11 define two methods \texttt{set-first!} and \texttt{set-second!} which are setters for these fields.
Line~13 defines a routine called \texttt{length} that will compute the length of a linked list by calling \texttt{length} on the \texttt{cdr} field as long as it is also a pair\footnote{Note that the \texttt{type-of} invocation in \cref{lst:classes} Line~15 returns a symbol that represents the name of the class. This is because classes are not reified as objects in our language (such as in SmallTalk~\cite[Chapter~5]{DBLP:books/aw/GoldbergR83}). They cannot be referenced directly \textit{except} via the \texttt{new} special form to create an instance.}.
We can enforce the properties of routines as follows.

\begin{lstlisting}[language=Stella, caption={The Pair class which has 2 kinds of operations, methods and routines.},label=lst:classes, float]
(class Pair
  (def-fields car cdr)
  
  (def-constructor (initialize-with initial-car initial-cdr)
    (set! car initial-car)
    (set! cdr initial-cdr))
  
  (def-routine (first) car)
  (def-routine (second) cdr)
  (def-method (set-first! new-car) (set! car new-car))
  (def-method (set-second! new-cdr) (set! cdr new-cdr))
  
  (def-routine (length) 
    (cond ((eq? cdr #undefined) 1)
          ((eq? (type-of cdr) |'Pair|) (+ 1 (length cdr)))
          (else 2))))
\end{lstlisting}

Routines that contain expressions with side-effects are rejected by the interpreter.
In our case they are \texttt{set!} and a couple of other special forms\footnote{The special forms forbidden in reactors and routines are \texttt{set!}, \texttt{spawn-actor}, \texttt{spawn-reactor}, \texttt{send}, \texttt{emit}, \texttt{monitor}, and \texttt{react-to}.}.
Together with property \#3, which is upheld using a run-time check, this ensures that routines will never have side-effects.
This must be checked at run-time because Stella is a dynamic language.
A run-time error occurs when a routine calls a method, for example \texttt{println} which performs IO.
	
Our current implementation of Stella uses \textit{size-change termination} (SCT) for higher-order programs~\cite{DBLP:conf/pldi/NguyenGTH19} to ensure
at run-time that routines terminate.
In a nutshell, this form of SCT dynamically constructs a \textit{size-change graph} based on the argument values of a routine call every time a routine is called.
Before entering a new routine call, the \textit{size-change termination principle} is used to compare the argument values to those of earlier calls to that routine (of the same class) higher on the call stack.
If the new argument list is not decreasing in size, a run-time exception is thrown.
Note that the algorithm assumes a well-founded partial order on values, for example, for numbers this is defined as \texttt{$|x| < |y|$}.
In this case a recursive routine such as \texttt{factorial} is permitted as long as recursion stops when the value of its argument has decreased to 0.

As an example, consider the excerpt of Stella code in \cref{lst:circular-list} that uses the \texttt{Pair} class of \cref{lst:classes}.
Here, a \texttt{Pair} is made to point to itself, creating a circular linked list. A subsequent call to \texttt{length} on the \texttt{Pair} gives
rise to a recursive call to \texttt{length} on the same class.
SCT rejects this call, since the argument values have not decreased since previous invocations.

\begin{lstlisting}[language=Stella, caption={Creating a circular data structure with \texttt{Pair} of \cref{lst:classes}.},label=lst:circular-list, float]
(def p1 (new Pair 'initialize-with 1 2))
(set-second! p1 p1)
(length p1) // successfully rejected via a run-time error
\end{lstlisting}

While \cite{DBLP:conf/pldi/NguyenGTH19} report that the overhead of their SCT analysis can be as high as a factor 100 (e.g.~for merge-sort), it is negligible for applications that perform a lot of work in between recursive calls (they give a factorial function as an example). 
In any case, overhead remains constant with respect to the size of the input of the program.
Crucially, it works for high-level programming languages such as Stella, and --- unlike other approaches such as specialized type systems --- SCT
requires no programmer assistance. Stella is not coupled to any specific SCT algorithm, or even to SCT itself. What \emph{is} part of the semantics of
Stella, is that long-lasting computations are illegal, and that termination of routines must be enforced by the Stella runtime.

\section{The Actor-Reactor Model}
\label{sec:actor-reactor-model}
With the base language defined, we can now describe how Stella tackles the problem described in \cref{subsec:coordination-of-paradigms}.
We sketched the general idea in \cref{subsec:solution-general-idea}.
The Actor-Reactor Model separates programs into \emph{actors} and \emph{reactors}.
Actors handle the parts of a program that involve long-lasting computations or side-effects, whereas reactors handle the parts that are inherently reactive, or that are more easily expressed using reactive programming.

Coordination between actors and reactors is achieved using \textit{data streams} that are produced by actors and reactors.
Stella defines a number of composition operators to statically and dynamically manage how data can flow to and from actors and reactors.
In the following sections we first introduce actors and data streams. 
In \cref{subsec:reactors} we introduce reactors and how to compose them with actors.

\subsection{Running Example: Wind Turbine Simulator}
\label{subsec:motivating-example-wind-turbine}
\begin{figure}[t]
	\centering
  	\includegraphics[width=0.7\linewidth]{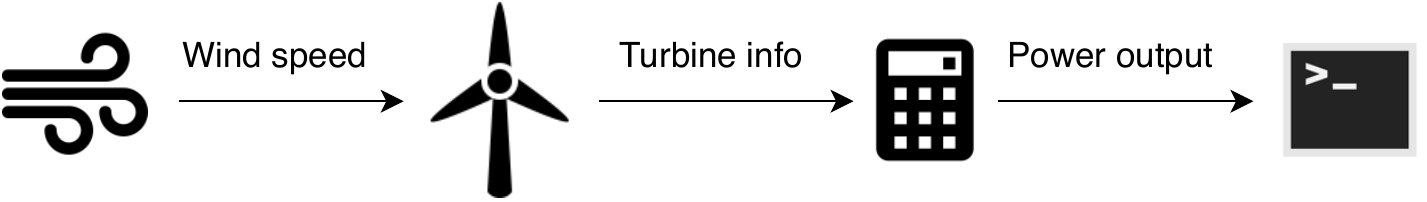}
  	\caption{Diagram of data flow in a wind turbine simulator consisting of a wind, a turbine, a turbine power calculator, and a console to print the result.}
  	\label{fig:turbine-data-flow}
\end{figure}

Consider a wind turbine simulator that calculates the real-time power output of a wind turbine.
A simple simulator can be defined using the 4 components depicted in \cref{fig:turbine-data-flow}.
From left to right: a blowing wind, a wind turbine, a power calculator, and a console to print the output.
Simulating wind can involve complex wind patterns and how they affect a turbine, or it can be a simple process that generates random values corresponding to the current wind speed.
Even the simple version is challenging because it involves a process ``a~wind'' that is continuously running and always changing.
It also requires reactive processes: A blowing wind impacts the rotation of the wind turbine, which then impacts the power output.
Important to note is that eliminating long lasting computations within reactive programs by itself is not enough to prevent them from blocking reactive programs.
It is equally important that reactive processes cannot be accidentally blocked by long lasting computations in other parts of the program, for example the simulation of wind patterns.

We model wind as an actor because it is inherently an active process that is always running and changing independently.
The console is also an actor because it performs IO.
We opt to model a turbine and the power calculator as reactors: a turbine \textit{reacts} to a wind, and a power calculator \textit{reacts} to changes in the turbine.

\subsection{Actors and Data Streams}
Actors are typically defined in terms of a \emph{behaviour} and a \emph{mailbox}~\cite{DBLP:conf/agere/KosterCM16}.
The behaviour of an actor describes its internal state and the messages that can be processed.
Messages that are sent to an actor are inserted into its mailbox (e.g.~a FIFO queue), and the actor continuously dequeues messages from its mailbox to process them one-by-one.
Actors in Stella are based on the Active Objects model~\cite{DBLP:conf/oopsla/YonezawaBS86, DBLP:conf/agere/KosterCM16, DBLP:journals/csur/BoerSHHRDJSKFY17}, where actor behaviours are defined similar to classes in object-oriented programming. 
In addition to receiving and processing messages, actors can be used to implement zero or more data streams to which they can \emph{emit} (publish) values.
The following sections explain the different parts of actor behaviours using the wind from our running example.
We explain how actor behaviours are defined, how data streams can be implemented using actors, and how actors can monitor data streams.

\begin{lstlisting}[language=Stella, caption={The \texttt{Wind} actor behaviour to implement a stream that represents wind speed.},label=lst:wind, float]
(actor Wind
  (def-stream speed 1)
  (def-fields rng)
  
  (def-constructor (init) (set! rng (new Random)))
    
  (def-method (blow) 
    (emit speed (integer-between rng 0 30))
    (sleep 10000)
    (send #self |'blow|)))
\end{lstlisting}

\subsubsection{Actor Behaviours}
\label{subsubsec:actor-behaviours}
\cref{lst:wind} defines the \texttt{Wind} actor behaviour.
An actor behaviour has a number of local fields, in this case 1~field called \texttt{rng} (Line~3).
A constructor called \texttt{init} is defined on Line~5, which initializes the \texttt{rng} field with a new random number generator (an object).
A method called \texttt{blow} without arguments is defined on Line~7, whose implementation we will explain later when we define data streams.
A \texttt{Wind} actor is thus capable of processing \texttt{blow} messages that are inserted into its mailbox, which amounts to invoking the corresponding \texttt{blow} method.

The special form \texttt{spawn-actor} is used to spawn new actors, in this case to create new winds. 
The following expression spawns an instance of \texttt{Wind}, which could be used to represent the mistral wind.
A reference is returned to the new actor, which is an object of type \texttt{ActorReference}.
To initialize the actor, \texttt{spawn-actor} takes the name of a constructor (as a symbol) and any arguments it requires.
Constructors are special kinds of methods that may only be invoked once, and only as the very first message an actor processes.
The act of spawning an actor via a constructor is semantically equivalent to spawning an actor and inserting a message in its empty mailbox that will initialize the actor.

\begin{lstlisting}[language=Stella]
(def mistral (spawn-actor Wind |'init|))
\end{lstlisting}

Actors communicate via asynchronous message passing via the \texttt{send} special form that inserts a new message in the mailbox of the designated actor. 
The following expression sends a \texttt{blow} message to \texttt{mistral}, which is expected to be an actor reference.
While in this case \texttt{blow} does not expect any arguments, they would be provided after the \texttt{'blow} symbol.
The message payload between actors (and reactors) is always passed by (deep) copy, and actors can send messages to themselves by sending them to \texttt{\#self}, which represents a reference to the current actor.

\begin{lstlisting}[language=Stella]
(send mistral |'blow|)
\end{lstlisting}

\subsubsection{Declaring Data Streams}
Every actor can export data streams.
The behaviour of an actor determines which data streams an actor implements via \texttt{def-stream} as seen on Line~2 of \cref{lst:wind}. 
This expression takes two arguments: the name of the stream and its \textit{arity}.
The name is used to uniquely identify a particular stream that belongs to a given actor, and the arity of a stream specifies the number of elements that must be emitted to the stream in one ``emit step''.
In our example a \texttt{Wind} actor exports a single stream called \texttt{speed} with arity~1.

Stream arity is used to emit new values that are intrinsically connected, and must therefore always change simultaneously.
For example, an actor might export a stream called \texttt{location} of arity 2 that represents coordinates in the form of \texttt{latitude} and \texttt{longitude}. 
Since they describe the real-time location of some real-world moving object, latitude and longitude should always change simultaneously.
Otherwise, a consumer of the stream might first update the entire application (e.g.~some world map) with a new value for latitude, and only after some time with the corresponding value for longitude.
After the first update the application would be in an inconsistent state, similar to a \textit{glitch} in reactive programming~\cite{DBLP:conf/esop/CooperK06}.
An alternative approach is to emit \texttt{location} as a single object. 
However, we will use stream arity to facilitate the composition of actors and reactors in \cref{subsec:reactors}.

\subsubsection{Publishing to Data Streams}
\label{subsubsec:publishing-to-data-streams}
Actors can emit values to their own data streams by using the \texttt{emit} special form.
Emitting a value amounts to sending the new value to all subscribers of the stream.
Consider the \texttt{blow} method in \cref{lst:wind}.
Whenever a \texttt{Wind} actor processes a \texttt{blow} message, on Line~8 the actor will \texttt{emit} a new value to the \texttt{speed} stream, which in this case is a random number between 0 and 30 representing the wind speed in meters/second.
Now, a special type of message (a publication) is added to the mailbox of subscribers, which are other actors or reactors.
Because the \texttt{speed} stream is defined with arity 1, \texttt{emit} only requires 1 argument.
The actor then sleeps for 10000ms (Line~9), and afterwards sends itself a new \texttt{blow} message to emit another value (Line~10).

\subsubsection{Qualifying and Monitoring Data Streams}
\label{subsubsec:monitoring-data-streams}
Two mechanisms are required to create a subscription on a data stream: \emph{qualification} and \emph{monitoring}.
Both are explained using the \texttt{Main} behaviour in \cref{lst:monitor} which can be seen as the console from our running example, but instead of monitoring and printing power output, it monitors and prints the wind speed.
When the \texttt{start} constructor is executed, Line~3 first spawns an instance of the \texttt{Wind} behaviour, and Line~4 exemplifies both qualification and monitoring.

\begin{lstlisting}[language=Stella, caption={Monitoring data streams with actors},label=lst:monitor, float]
(actor Main
  (def-constructor (start) 
    (def sirocco (spawn-actor Wind |'init|))
    (monitor sirocco.speed |'print-wind|))
    
  (def-method (print-wind wind-speed) 
    (println "the new wind speed is: " wind-speed)))
\end{lstlisting}

Qualification is the act of designating a reference to a particular stream exported by a particular actor (or reactor).
On Line~4, the expression \texttt{sirocco.speed} evaluates to an object of class \texttt{Stream} that represents a reference to the \texttt{speed} stream exported by the \texttt{sirocco} actor.
Data will only start flowing once a consumer (an actor or reactor) subscribes to the stream, in which case data flows directly from the producer to the consumer.

Actors subscribe to data streams by monitoring them.
This is exemplified by Line~4, where the \texttt{sirocco.speed} stream is monitored for changes.
Whenever this stream emits a new value, a \texttt{'print-wind} message will enter the mailbox of the actor, which is processed by the corresponding \texttt{print-wind} method on Line~6.
This method requires exactly 1 formal parameter because the \texttt{speed} stream has an arity of 1.
We will see in \cref{subsec:reactors} that reactors do not explicitly monitor data streams like actors, and instead they will automatically react to values that are emitted to data streams.

\subsection{Reactors}
\label{subsec:reactors} 
Reactive languages rely on 2 fundamental mechanisms.
First, at compile-time, the program text is compiled to a DAG that consists of source nodes that represent the input of the reactive program, sink nodes represent the output of the reactive program, and internal nodes represent all computations that occur between the sources and sinks.
Second, at run-time, a \textit{reactive engine} is responsible for propagating new values through the DAG, such that the calculation that makes up the output remains consistent with the values of the input.
The reactive engine is smart enough to prevent glitches~\cite{DBLP:conf/esop/CooperK06}.

In the following sections we gradually explain Stella's reactors by implementing a simple reactive wind turbine and power calculator. 
Stella features 2 different ways to statically compose \textit{reactor behaviours}: via \emph{point-wise} and \emph{point-free} composition, named after function composition in Haskell~\cite[Chapter~5]{Lipovaca:2011:LYH:2018642}.
In \cref{subsubsection:reactors} we explain the run-time semantics of reactors and how they manage dependencies on data streams.

\subsubsection{Definitions}
A reactor consists of 3 layers.
\begin{description}
	\item [Layer 1: Reactor Behaviour] A \emph{reactor behaviour} describes the static properties of a reactive program, represented by a DAG that is constructed at ``DAG compile-time'' (a pre-processing step of our interpreter).
		The DAG is constructed from the program text, for example the reactor behaviour of \cref{lst:reactor-behaviour-basic} which we will explain in \cref{subsubsection:reactor-behaviours}.
		It describes the source nodes, sink nodes, and internal nodes of the reactive program, and all of the dependencies between them.
		We may refer to an actor or reactor behaviour as simply ``behaviour'' if it is clear from context to which one we are referring.

	\item [Layer 2: Reactor Deployment] Reactive languages usually store the run-time information of a reactive program (e.g.~node values and local state) directly in the nodes of the DAG.
		However, in our case reactor behaviours can be composed and reused by multiple reactors, and every use of a DAG can be in a different state depending on the values that were propagated.
		Therefore, a \emph{reactor deployment} represents a specific instance of a reactor behaviour.
		In other words, a reactor deployment stores all run-time information pertinent to a specific instance of a DAG that is used by a specific reactor.

	\item [Layer 3: Reactor]
		A \emph{reactor} is process with a mailbox and a \emph{vat} (collection) of reactor deployments.
		It is the driving force behind a reactive program: a reactor continuously dequeues values from its mailbox and propagates them through the destined deployment via a built-in reactive engine.
		Similar to how actors are spawned from actor behaviours, a reactor is spawned from a reactor behaviour (that represents a DAG).
		At spawn-time, the reactor creates an initial deployment for this DAG, which we call the \emph{root deployment}. 
		A reactor has exactly 1 output stream of arity \texttt{n}, where \texttt{n} corresponds to the number of sinks of the root deployment.
		Every time the root deployment updates, its sink values are emitted on the output stream of the reactor.
		Our definition of reactors intentionally covers deployments that give rise to other deployments within the same vat, but we will not discuss those features in this paper.
		Reactors will therefore always contain exactly 1 deployment (namely the root deployment).

\end{description}

\subsubsection{Basic Reactor Behaviours}
\label{subsubsection:reactor-behaviours}
A reactor behaviour is the textual representation of the DAG of a reactive program: it has a name, at least 1 source, at least 1 sink, and any number of internal nodes that describe the computations between the sources and sinks.
An example of a computation is the theoretical power output of a wind turbine (in Watt) which is based on the area swept by its blades, the air density surrounding the turbine, and the velocity of the wind~\cite{windturbinespower}.
It can be calculated as follows:  
\[ 
Power\:(W) = 0.5 \times Swept\:Area\:(m^{2}) \times Air\:Density\:(kg/m^{3}) \times Velocity^{3}\:(m/s)
 \]

\begin{figure}[t]
\centering
\subcaptionbox{\texttt{WindPower}\label{fig:theoretical-wind-power-graph}}{\includegraphics[width=.45\linewidth]{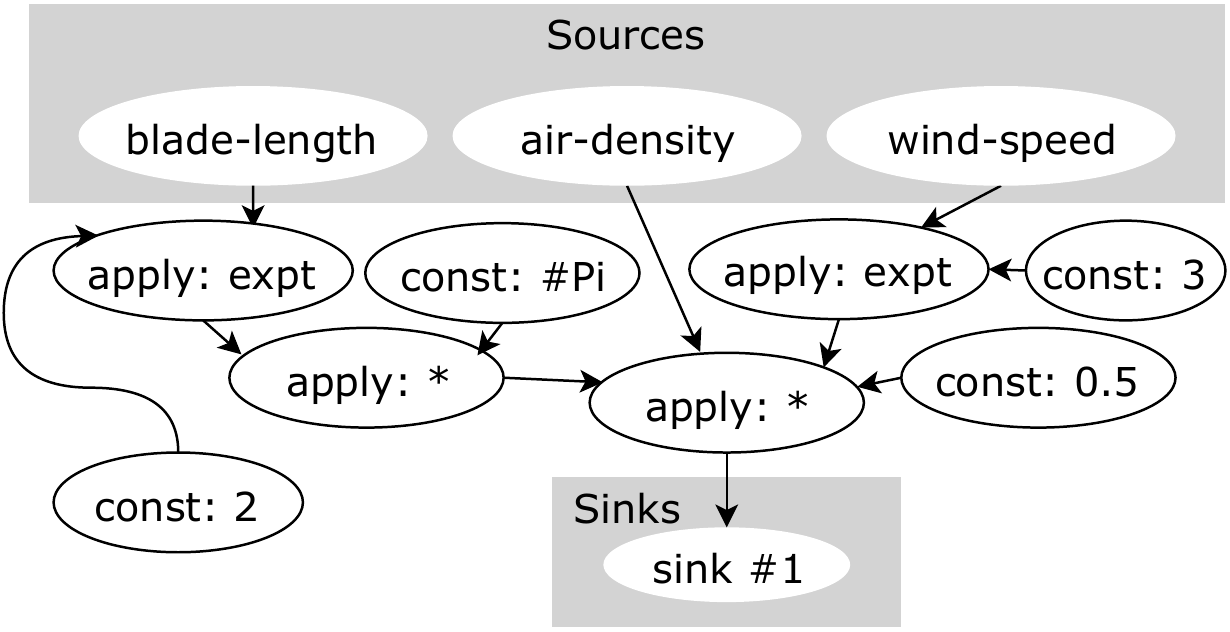}}
\subcaptionbox{\texttt{PowerOutput}\label{fig:power-output-graph}}{\includegraphics[width=.45\linewidth]{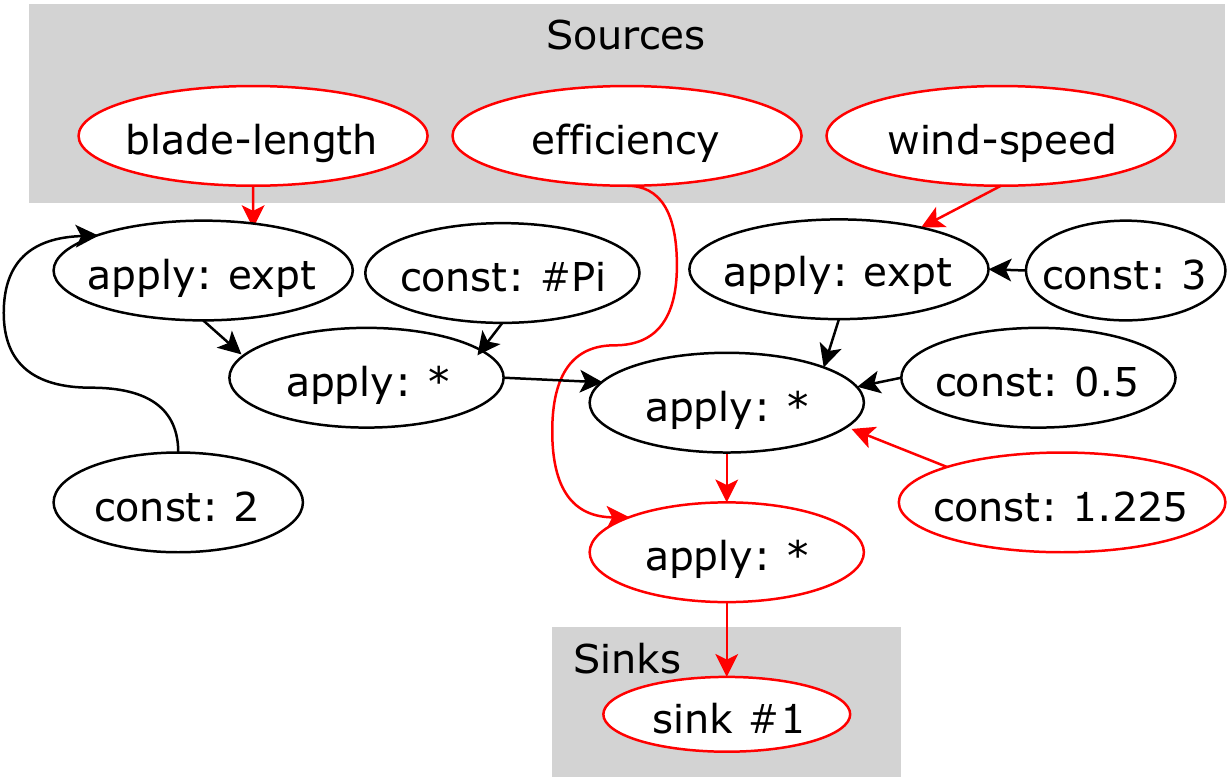}}

\caption{Side-by-side comparison for the DAGs of \texttt{WindPower} and \texttt{PowerOutput} of \cref{lst:reactor-behaviour-basic} and~\ref{lst:reactor-behaviour-pointwise}. Nodes and dependencies introduced by \texttt{PowerOutput} are highlighted in red.}
\end{figure}

This formula is implemented in \cref{lst:reactor-behaviour-basic} in a reactor behaviour called \texttt{WindPower} with 3 source nodes and 1 sink node.
Its 3 sources are called \texttt{blade-length}, \texttt{air-density}, and \texttt{wind-speed}.
There is one local variable called \texttt{swept-area}, and one sink node that is immediately linked to the result of a multiplication.
The DAG of this behaviour is depicted in \cref{fig:theoretical-wind-power-graph}.
Every invocation of a routine is depicted by an ``apply'' node, and all expressions without dependencies are wrapped in a ``const'' node that will be computed when the DAG is compiled.
Since the run-time values that are propagated through the DAG are regular objects, reactors can only invoke routines on those objects, and the invocation of regular methods will result in a run-time error.

\begin{lstlisting}[language=Stella, caption={The \texttt{WindPower} behaviour calculates the maximum theoretical power output of a turbine.},label=lst:reactor-behaviour-basic, float]
(reactor (WindPower blade-length air-density wind-speed)
  (def swept-area (* #Pi (expt blade-length 2)))
  (out (* 0.5 swept-area air-density (expt wind-speed 3))))
\end{lstlisting}

\subsubsection{Point-wise Graph Composition}
\label{subsubsec:pointwise-graph-composition}

\begin{lstlisting}[language=Stella, caption={Point-wise composition of reactor behaviours.},label=lst:reactor-behaviour-pointwise, float]
(reactor (PowerOutput blade-length efficiency wind-speed)
  (def wind-power (tick WindPower blade-length 1.225 wind-speed))
  (out (* efficiency wind-power)))
\end{lstlisting}

While the \texttt{WindPower} behaviour computes the theoretical power output of a turbine, a more accurate calculation takes into account turbine efficiency (typically between 10--30\%~\cite{windturbinespowerefficiency}).
To this end, \cref{lst:reactor-behaviour-pointwise} defines a behaviour called \texttt{PowerOutput} that shows how reactor behaviours can be composed in a point-wise manner.
It has 3 sources called \texttt{blade-length}, \texttt{efficiency} and \texttt{wind-speed}.
They correspond to the 3 pieces of information that a wind turbine is expected to produce to be able to calculate its power output. 
Line~2 performs a point-wise composition of reactor behaviours via the \texttt{tick} special form.
This can be thought of as a function application but for reactor behaviours, of which the result is bound to the \texttt{wind-power} variable.
Line~3 then scales the theoretical output by the efficiency of the turbine.

A \texttt{tick} operation is resolved at compile-time as the inlining of a DAG.
First, the source nodes of the composed behaviour (\texttt{WindPower}) are connected to the corresponding arguments of \texttt{tick}.
Second, the sink node of the composed behaviour is connected back into the composer, in this case to all nodes in \texttt{PowerOutput} that use the \texttt{wind-power} variable (multiple sinks would be defined via a \texttt{def-values} special form).
The resulting graph is depicted in \cref{fig:power-output-graph}, where the nodes and dependencies introduced by \texttt{PowerOutput} are highlighted in red.
For brevity, in the \texttt{tick} expression we assume a default value of 1.225kg/m$^{3}$ (the air density at $15^\circ$C at sea level).

While it is not part of our goals, the structure of the DAG can be optimised when it is compiled. 
In this case the source and sink nodes of the inlined \texttt{WindPower} behaviour were automatically eliminated because they are redundant, and because (by definition) sources and sinks can only exist at the boundaries of a DAG.
For example, the constant value 1.225 provided by the composer in \cref{lst:reactor-behaviour-pointwise} is represented by a constant node in \cref{fig:power-output-graph} (depicted in the bottom right). 
This node is directly connected to the apply node because the original \texttt{air-density} source node (which it replaces) was eliminated.

\subsubsection{Behaviour Stream Composition}
From the \texttt{PowerOutput} behaviour we know that a wind turbine should provide its blade length, efficiency, and the current wind speed affecting it. 
The simplest possible implementation of a turbine that we can think of is defined in \cref{lst:turbine}. 
This reactor behaviour has 3 sources: \texttt{blade-length} is a number usually between 20 and 80 (meters), \texttt{efficiency} is a number between 0.1 and 0.3 (10-30\%), and \texttt{wind} should be a reference to an actor.
The output of \texttt{Turbine} is a stream of arity 3 that echoes the blade length and efficiency, and the qualification ``\texttt{wind.speed}'' will echo the contents of the \texttt{speed} stream.

\begin{lstlisting}[language=Stella, caption={The \texttt{Turbine} behaviour implements a simple wind turbine.},label=lst:turbine, float]
(reactor (Turbine blade-length efficiency wind)
  (out blade-length efficiency wind.speed))
\end{lstlisting}

A qualification in the body of a reactor behaviour represents a dependency to a stream.
However, handling such dependencies can be quite tricky since there are potentially two kinds of changing values.
First, the exporting actor of a stream may change, for instance when a new actor reference is propagated for the \texttt{wind} source node (e.g. \texttt{sirocco} or \texttt{mistral}).
Second, the \texttt{speed} stream is continuously emitting new values. 
A reader may recognise this as a \textit{higher-order} stream~\cite{DBLP:conf/oopsla/MeyerovichGBCGBK09}.
The source node is conceptually a stream of values, and every value is an actor (or reactor) that exports streams.

\begin{figure}[t]
	\centering
  	\includegraphics[width=0.7\linewidth]{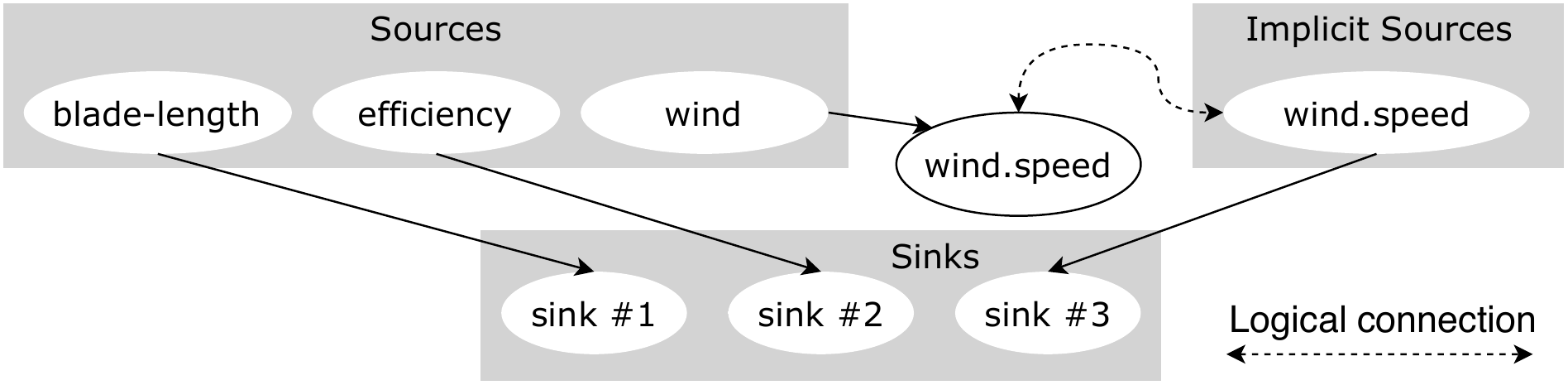}
  	\caption{DAG of the \texttt{Turbine} behaviour.}
  	\label{fig:implicit-sources}
\end{figure}

Inspired by the compilation of the \textit{async} expression in Elm~\cite{DBLP:conf/pldi/CzaplickiC13}, every qualification is compiled to 2 graph nodes. 
First, an internal node manages the dependency on the referenced stream.
Second, an \emph{implicit source node} is responsible for processing the values emitted by the stream.
The resulting graph for \texttt{Turbine} is depicted in \cref{fig:implicit-sources}. 
When the source node \texttt{wind} changes to a different actor (e.g.~another wind) then this new value is propagated to the special qualification node.
This node unsubscribes from its current stream (if present) and subscribes to the newly referenced \texttt{speed} stream. 
The value of the implicit source is immediately changed to the most recent value emitted by the newly referenced stream.
From that point onwards, whenever publications of the new stream enter the mailbox of the reactor, the reactor will process them by changing the value of the corresponding implicit source node.

\subsubsection{Point-free Graph Composition}
\label{subsubsec:pointfree-behaviour-composition}
There are two ways to implement a wind turbine that is linked to a power calculator. 
Either a new reactor is spawned for each of them and their input/output streams are subsequently linked together, or the 2 reactor behaviours are composed and spawned as a single reactor.
Both approaches are valid, and which one is more desirable depends on the application.
We take the second approach by composing the 2 behaviours via \emph{point-free} graph composition.

In Haskell, new functions can be defined point-free via a function composition operator.
The composition \texttt{f $\circ$ g} (``\texttt{f} after \texttt{g}'') is a function that first applies \texttt{g} to its argument, then \texttt{f} to the value returned by \texttt{g}; \texttt{(f . g) x = f(g(x))}. 
Similarly, the \texttt{ror} operator composes reactor behaviours: \texttt{r$_{1}\:\circ$ r$_2$} constructs a new behaviour where data is first propagated through \texttt{r$_2$} and then through \texttt{r$_1$}. 
The DAG of this behaviour is the composition of \texttt{r$_1$} and \texttt{r$_2$}, where the sinks of \texttt{r$_2$} are connected to the sources of \texttt{r$_1$}.

As an example, \cref{lst:reactor-behaviour-pointfree} defines a new behaviour called \texttt{TurbinePowerOutput} that combines the behaviours of \texttt{Turbine} and \texttt{PowerOutput}.
We designed those behaviours such that they can be easily composed, i.e. the sinks of \texttt{Turbine} directly match the sources of \texttt{PowerOutput}.
If the behaviours would not directly fit together, intermediate behaviours can take care of reordering sources and sinks, or transforming data.

\begin{lstlisting}[language=Stella, caption={Point-free composition of reactor behaviours.},label=lst:reactor-behaviour-pointfree, float]
(reactor TurbinePowerOutput (ror PowerOutput Turbine))
\end{lstlisting}

The \texttt{ror} operator is capable of connecting \textit{multiple} ``input'' behaviours to 1 ``output'' behaviour.
The following expression defines a new behaviour \texttt{R} that is the composition of an output behaviour $R_{out}$ with input behaviours $R_1$ to $R_n$.
\begin{lstlisting}[language=Stella, mathescape=true]
(reactor R (ror R$_{out}$ R$_1$ R$_2$ $\dots$ R$_n$))
\end{lstlisting}

Behaviour \texttt{R} can be compiled as long as the number of sinks of all input behaviours matches the number of sources in $R_{out}$. 
If this is the case, they will be connected in order from left to right to construct the behaviour \texttt{R}.
The sources of \texttt{R} will be the same as the sources of the inputs, ordered from left to right.
\[
sources(R) \: := \: sources(R_1) \: + \: sources(R_2) \: + \: \dots \: + \: sources(R_n)
\]
The sinks of \texttt{R} are the same as the sinks of $R_{out}$.
\[
sinks(R) \: := \: sinks(R_{out})
\]

Additional point-free graph composition operators with different semantics are conceivable, such as the \texttt{parallel} and \texttt{parallel*} operators in~\cite{DBLP:conf/oopsla/OeyenRVM18}.

\subsubsection{Run-time Semantics of Reactors: Spawning and Linking}
\label{subsubsection:reactors}
We now complete the running example of \cref{subsec:motivating-example-wind-turbine} by showing how the program is started, and how actors and reactors are linked together.
We will focus on the composition of actors and reactors rather than the internal semantics of reactors.
Internally, it suffices to know that every reactor has a reactive engine that is responsible for propagating values from sources to sinks.
Similar to Flapjax~\cite{DBLP:conf/oopsla/MeyerovichGBCGBK09} and REScala~\cite{DBLP:conf/aosd/SalvaneschiHM14}, we based our propagation algorithm on that of FrTime~\cite{DBLP:conf/esop/CooperK06}, but without the complexity of a dynamic dependency graph.
It ensures that only the parts of a DAG that are affected by a change will be recomputed, and that computations produce no glitches when multiple nodes change simultaneously.

\cref{lst:turbine-simulator-main} implements the \texttt{Main} program for the running example.
Its purpose is to spawn an actor representing a wind, spawn a reactor representing a wind turbine and its power output calculation, and to print this power output to the console. 
This functionality is implemented by the \texttt{start} constructor. 
Most of the expressions have been discussed previously.
Line~3 spawns an instance of the \texttt{Wind} actor behaviour and Line~4 sends it a \texttt{blow} message.
The wind will now start periodically emitting values.

The \texttt{spawn-reactor} expression on Line~5 spawns a new reactor that is now waiting for data on its sources. 
Remember that a reactor is defined as a vat of reactor deployments, and in this case \texttt{TurbinePowerOutput} will be the root deployment, as well as the only deployment for this reactor throughout its lifetime.

Reactors are linked to actors (or other reactors) via the \texttt{react-to} special form that will change the values of source nodes.
If the new value of a source node is an object of type \texttt{Stream}, instead of changing the value of the source node, the reactor will automatically create a subscription on the stream (possibly replacing an existing subscription).
Then, whenever new values are emitted by this stream, they enter the mailbox of the reactor which will process them by modifying the value of the corresponding source node. 
In \cref{lst:turbine-simulator-main} the sources of \texttt{turbine} are changed on Line~6 to 80, 0.3, and \texttt{sirocco}.
In the context of our application, this means that this is a reactor that represents a turbine with a blade length of 80 meters, an efficiency of 30\%, which is influenced by the \texttt{sirocco} wind.
To print the output of the turbine the console, on Line~7 the main actor monitors the turbine for changes. 
Reactors export exactly 1 output stream called \texttt{out}.

Reactors are aware of stream arity, and the \texttt{react-to} composition operator can be used to react to streams with an arity greater than 1.
In this case, the reactor requires exactly as many source nodes as the arity of the input streams.
For example, a reactor that is made to react (via \texttt{react-to}) to a stream of arity 2 will require exactly 2 source nodes.
Whenever the input stream emits new values they will enter the mailbox as a single publication, but (in this case) the value of the 2 source nodes of the reactor will change simultaneously.
The \texttt{react-to} composition operator ensures that there is a one-to-one mapping between its arguments and the source nodes of the reactor.

\begin{lstlisting}[language=Stella, caption={The \texttt{Main} program for the wind turbine simulator of \cref{subsec:motivating-example-wind-turbine}.},label=lst:turbine-simulator-main, float]
(actor Main 
  (def-constructor (start)
    (def sirocco (spawn-actor Wind |'init|))
    (send sirocco |'blow|)    
    (def turbine (spawn-reactor TurbinePowerOutput))
    (react-to turbine 80 0.3 sirocco)
    (monitor turbine.out |'print|))
    
  (def-method (print watt)
    (println "turbine produced: " (round (/ watt 1000000)) " MW")))
\end{lstlisting}

\section{Evaluating the Awkward Squad for Reactive Programming}
We introduced the awkward squad for reactive programming as a set of issues that are essential for real-world software development, but that do not fit within reactive programming.
In this section we investigate the extent to which these issues can be present in existing reactive languages and frameworks.
We find that it is indeed the case that existing languages and frameworks expose operations or mechanisms to developers that fall within one or all issues of the awkward squad.
In many cases these operations and mechanisms will be unavoidable for developers, either because they are an inherent part of the language or framework, or because otherwise it would be impossible to write certain applications, e.g.~applications with a GUI.
In this paper we do not investigate the extent to which possibly issue-causing operations are used in real applications, and if they are present, which bugs they can possibly cause in those specific applications.

\cref{table:evaluation} lists a number of reactive languages and frameworks that we consider to be representative for the state-of-the-art.
They are FrTime~\cite{DBLP:conf/esop/CooperK06} (Racket), Flapjax~\cite{DBLP:conf/oopsla/MeyerovichGBCGBK09} (JavaScript), REScala~\cite{DBLP:conf/aosd/SalvaneschiHM14} (Scala), ReactJS~\cite{reactjs} (JavaScript), Akka Streams~\cite[Chapter~13]{roestenburg2015akka} (Scala), and RxJS~\cite{rxjs} (JavaScript).
We used them to implement our running example of \cref{subsec:motivating-example-wind-turbine}\footnote{All code from this paper and the implementations used to guide our evaluation of the different languages and frameworks are available in an artefact. To download this artefact, see the supplementary material on the first page of this paper.}.
Below the double horizontal line we list 3 other reactive languages which we did not use to implement the application (for technical reasons) but which we classified according to their respective papers.
They will be discussed in \cref{subsec:evaluation-additional-mentions}.
Based on our findings we categorised these languages and frameworks as follows. 

\begin{description}
	\item [Reactive Thread Hijacking Problem (RTHP)] Does the language or framework solve the Reactive Thread Hijacking Problem?
		In other words, does the language or framework prevent infinite computations from blocking the reactive program?
		We make no distinction between eventual reactivity and strong reactivity.
		If not, we discuss the features that can be used to block the reactive program.
		
	\item [Reactive Update Order Leak (RUOL)] Does the language or framework solve the Reactive Update Order Leak?
		In other words, does the language or framework disallow side-effects in internal nodes of the DAG? 
		If not, we will highlight the features where side-effects can be executed inside the DAG.

	\item [Reactive/Imperative Impedance Mismatch (RIIM)] 
		In \cref{subsec:coordination-of-paradigms} we listed a number of different mechanisms used by reactive languages to be able to embed reactive code within imperative code.
		We discuss, to the best of our knowledge, which of the listed mechanisms are used.
\end{description}

\begin{table}[tb]
\centering
\begin{tabularx}{\linewidth}{ n|o|o|X }

  & \textbf{RTHP} & \textbf{RUOL} & \textbf{RIIM} \\ \hline 
  \textbf{FrTime} & $\times$ & $\times$ & Periodic polling, hidden concurrency, domain specific features, metaprogramming \\ \hline
  \textbf{Flapjax} & $\times$ & $\times$ & Metaprogramming, domain specific features \\ \hline
  \textbf{REScala} & $\times$ & $\times$ & Metaprogramming \\ \hline
  \textbf{ReactJS} & $\times$ & $\times$ &  Metaprogramming, domain specific features \\ \hline
  \textbf{Akka Streams} & $\times$ & $\times$ & \checkmark \\ \hline 
  \textbf{RxJS} & $\times$ & $\times$ & Metaprogramming \\ \hline
  \textbf{Stella} & \checkmark & \checkmark & \checkmark \\ \hhline{=|=|=|=}
  \textbf{Elm} & $\times$ & $\times$ & Domain specific features, special forms \\ \hline
  \textbf{ActiveSheets} & \checkmark & \checkmark & \checkmark \\ \hline
  \textbf{Coherence} & $\times$ & \checkmark &  \checkmark \\ \hline
\end{tabularx}
\caption{Categorisation of reactive languages and frameworks.}
\label{table:evaluation}
\end{table}

\subsection{FrTime}
FrTime is a functional reactive programming language built in Racket that can be interacted with via the Racket Read-Eval-Print Loop~\cite{DBLP:conf/esop/CooperK06}.

\begin{description}
	\item [RTHP] The execution time of expressions in FrTime is unrestricted.
	While there exists a limited set of built-in functions (e.g.~arithmetic) that always terminate, the built-in \texttt{lift-strict} primitive is used to integrate any (possibly infinitely looping) Racket function with the DAG.
	The implementation of FrTime exposes multiple threads of execution to developers: one thread is responsible for updating the reactive program, and another thread is responsible for the REPL.
	To implement an infinite loop that represents a wind, we blocked one thread of execution to continuously modify a source node of the dependency graph.
	
	\item [RUOL] Any Racket function may be used in internal nodes of the DAG via the aforementioned \texttt{lift-strict} function.
		Additionally, FrTime offers abstractions for ``event streams'' that can be mapped and filtered (via \texttt{map-e} and \texttt{filter-e} respectively) using using regular Racket functions.
	
	\item [RIIM]
		We have identified 4 mechanisms that are used to embed reactive code within imperative code.
		Firstly, there are 2 built-in signals called \texttt{seconds} and \texttt{milliseconds} that are updated automagically by the runtime, which can be used for periodic polling.
		Secondly, multiple threads of execution are exposed to developers.
		While the Racket REPL thread can be used to send and receive values to and from the reactive program, the semantics of their interaction is not part of the language definition.
		Thirdly, FrTime has domain specific features in the form of a wrapper around the Racket GUI toolkit~\cite{DBLP:conf/flops/IgnatoffCK06}, which automatically integrates with the DAG.
		Lastly, source nodes of the DAG can be manually defined via \texttt{cells} and \texttt{event-receivers}, and they can be assigned to via \texttt{set-cell!} and \texttt{send-event} respectively.
		The semantics of modifying a source node is unspecified, especially when assignments to the same sources occur in multiple threads.
\end{description}

\subsection{Flapjax}
Flapjax is a reactive programming language based on JavaScript~\cite{DBLP:conf/oopsla/MeyerovichGBCGBK09}.

\begin{description}
	\item [RTHP] Nodes in the dependency graph consist of arbitrary JavaScript functions, which are unrestricted in their execution time.
	To create an infinite loop that implements a wind without blocking the browser, we implemented a non-blocking loop by wrapping JavaScript's \texttt{setTimeout} to asynchronously schedule an event in the JavaScript event-loop. 
	Some built-in operations can unintentionally block the reactive program.
	For example, the \texttt{evalForeignScriptValE} operation is applied to a reactive value that contains a URL.
	It retrieves and evaluates the (foreign) JavaScript file on the URL, and publishes its return value as a new reactive value~\cite{flapjax-docs}. 
	
	\item [RUOL] Flapjax programs consist of arbitrary JavaScript expressions that may involve arbitrary side-effects inside the DAG.
		Some built-in operations execute side-effects in the DAG, such as the GUI modification operators \texttt{insertDomB} and \texttt{insertDomE} to insert a reactive value in the browser DOM, operations such as \texttt{getWebServiceObjectE} to perform XMLHttpRequests, or the aforementioned \texttt{evalForeignScriptValE} to evaluate an arbitrary JavaScript file.
	
	\item [RIIM] To embed reactive code within imperative JavaScript code, Flapjax has 2 mechanisms.
		Source nodes of the DAG can be manually created from the GUI via operations such as \texttt{extractValueB} and \texttt{extractValueE} which are updated automatically by the runtime, and ex nihilo via \texttt{receiverE} (to create a new event stream) and \texttt{sendEvent} (to modify an event stream).
		Flapjax also offers special features to construct GUI elements that automatically integrate with reactive values. 
\end{description}

\subsection{REScala}
REScala is a reactive programming library in Scala that unifies the concepts of functional reactive programming with object-oriented programming~\cite{DBLP:conf/aosd/SalvaneschiHM14}.

\begin{description}
	\item [RTHP]
		REScala imposes no restrictions on the execution time of expressions inside the DAG of a reactive program.
		Since it is built as a library, regular Scala functions are used to perform computations on reactive values.
		To model a wind from our running example without blocking the reactive program, we manually constructed a new Scala thread with an infinite loop that non-reactively modifies a source node of the reactive program.

	\item [RUOL]
		Since REScala is conceived as a library, the Scala functions used in nodes of the DAG may perform arbitrary side-effects.
		As a design guideline, the REScala documentation explicitly mentions that functions inside the DAG must be pure~\cite[1.5.3]{rescala-side-effects}.
	
	\item [RIIM]
		To embed reactive code within imperative code, REScala offers special features to manually create new source nodes of the DAG (\texttt{Var}s) and to modify them via assignment.
		Conversely, callbacks can be installed on sink nodes of the DAG to act on their changes.
		
\end{description}

\subsection{ReactJS}
ReactJS is a JavaScript reactive GUI framework developed by Facebook~\cite{reactjs}, which is used to develop reactive web applications and mobile applications~\cite{reactnative}.

\begin{description}
	\item [RTHP]
		The types of expressions that constitute a ReactJS program are regular JavaScript expressions, and are unrestricted in their computation time.
		To implement a wind without blocking the browser, we used JavaScript's \texttt{setInterval} that calls a function on a repeated interval.

	\item [RUOL]
		There are no restrictions on side-effects inside a ReactJS dependency graph.
		The documentation mentions that reactive components should be pure only with respect to their ``props'' object, which is a framework-provided object that is used to create dependencies between reactive components~\cite{reactjs-props-side-effects}.
		
	\item [RIIM]
		The state of a reactive component is manually modified via a special \texttt{setState} method that triggers a new propagation cycle.
		ReactJS offers domain specific features (via \textit{JSX templates}) to construct user interfaces that automatically display the values of sink nodes of the dependency graph.
\end{description}

\subsection{Akka Streams}
Akka Streams is a streaming library in Scala based on the Akka actor library~\cite[Chapter~13]{roestenburg2015akka}.
We had some difficulties reproducing the semantics of our running example in Akka Streams because we found it to be very difficult to create dependency graphs that are not linear (multiple sources or sinks) and which have similar update semantics.

\begin{description}
	\item [RTHP]
		There are no restrictions on performing long lasting computations inside the DAG, e.g.~via the stream operator \texttt{map} to apply a regular Scala function to a stream, or \texttt{zipWith} to combine 2 or more streams via a regular Scala function.

	\item [RUOL]
		Many built-in streaming operators are designed to be without side-effects.
		While we could not find explicit programmer guidelines discussing side-effects in the Akka Streams documentation, we believe that stream operators are intended to be pure, including those that accept arbitrary Scala functions (e.g.~\texttt{map} or \texttt{zipWith}). 
		We found at least some interest by users of the Akka project on GitHub for stream operators that execute side-effects. 
		In one instance a user requested a novel operator to execute side-effects without performing a value transformation, noting that the only way to achieve the desired effect was via the \texttt{map} operator (which, in the experience of the user, lead to code duplication)~\cite{akkagithub1}. 
		In response to this issue a new \texttt{wireTap} stream operator was added to Akka version 2.5.13.
		A new variant of this operator (with different semantics) is currently being requested by a different user~\cite{akkagithub2}.
		This anecdotal evidence is at least an indication that functional purity is not always upheld by the users of Akka.
		Some operators with side-effects are built-in, for example, the \texttt{log} operator logs the elements flowing through a stream, and the \texttt{ask} operator sends an asynchronous message to an actor.
		Sink nodes also perform side-effects to act on the elements of a data stream, e.g.~to print results to the console via a \texttt{forEach} operator.
		
	\item [RIIM]
		As far as we know, Akka Streams has a clean separation between the code that is responsible for supplying values to the reactive program (which are actors) and the code that is responsible for the reactive program itself (which are actors that run data streams).
		Note that this is not always the case, since there exist many operators to create new source nodes ex nihilo whereby the reactive program (an actor running data streams) is itself responsible for collecting/retrieving its input data.
		For example, the \texttt{FiloIO.fromPath} operator that creates a source node to read the contents of a file.
\end{description}

\subsection{RxJS}
RxJS~\cite{rxjs} is a streaming library for JavaScript based on the ReactiveX specification~\cite{reactivex}, which has currently been implemented in 18 languages.

\begin{description}
	\item [RTHP]
		There are no restrictions on the execution time of the expressions that constitute a stream, and long lasting computations will block the entire program.
		Implementations of ReactiveX in other languages may support \textit{Schedulers} which are designed to introduce multithreaded processing to streams, but this can cause other issues, especially when used in combination with side-effects.
	
	\item [RUOL]
		Many built-in streaming operators are designed to be without side-effects, and the documentation of RxJS describes operators in general as ``pure functions''~\cite{reactivexoperators}.
		We believe that operators that rely on arbitrary JavaScript functions (e.g.~\texttt{map} or \texttt{zipWith}) are intended to be functionally pure as well (but which cannot be enforced).
		In some cases side-effects are unavoidable, for example when defining a type of source node called a \texttt{Subject}, which is manually updated to start the propagation of new values. 
		Frameworks in the family of ReactiveX (such as RxJS) also offer a whole range of ``\texttt{Do}'' operators that are specifically designed to execute side-effects within a reactive program without performing a value transformation~\cite{reactivexdo}. 
		
	\item [RIIM]
		Programmers in RxJS can manually create and update new source nodes (a \texttt{Subject}) of the DAG, and they can manually create new streaming operators using metaprogramming to read from unsupported data sources. 
		Callbacks are registered on sink nodes of the program to imperatively act upon their changes (e.g.~by modifying the GUI, or printing to the console). 
\end{description}

\subsection{Stella}

\begin{description}
	\item [RTHP]	 Stella solves the Reactive Thread Hijacking Problem by eliminating infinite computations from reactive programs.
		Whether our implementation also solves the problem of responsiveness in general depends on the expectations of the application developer. 
	We believe there is no one-size-fits-all solution to ensure that an application remains ``reactive'' or ``responsive'', since their meaning is likely to change depending on the application requirements or domain.
	The Actor-Reactor Model facilitates restricting certain parts of the application (the reactive parts) to provide extra guarantees with respect to responsiveness or computational complexity without introducing the problems that we identify in this paper. 
	The design choice that we made in Stella is to enforce eventual reactivity via size-change termination. 
	Thus, Stella ensures that reactors must \textit{eventually} terminate, and that the execution thread of a reactor is not accidentally hijacked by computations in other parts of the program (other actors or reactors).
	In different application domains with stronger memory or timing requirements (e.g.~safety systems, robotics, \dots) it would be conceivable to further restrict reactors, for example by imposing strong reactivity and bounded-size mailboxes.

	\item [RUOL] Stella solves the Reactive Update Order Leak by ensuring that effectful computations cannot be part of the dependency graph of a reactive program.
	This is realised by routines in the object-oriented base language.
	However, using methods and routines may be a burden for programmers, since they must make an effort to correctly program a piece of functionality as a regular method or as a routine.
	When using built-in classes or libraries they must also know whether functionality is offered as a method or as a routine.

	\item [RIIM] Stella solves the Reactive/Imperative Impedance Mismatch.
	It ensures that the embedding of reactive code within imperative code does not introduce the Reactive Thread Hijacking Problem and the Reactive Update Order Leak, and that the semantics of embedded reactive code are clear. 
	There are 2 composition operators available to actors: \texttt{react-to} and \texttt{monitor}.
	To imperatively change the source nodes of a reactive program, actors must use \texttt{react-to} with clearly defined semantics: a message is sent to the reactor that, when processed, changes its source nodes.
	To imperatively act on the changes of reactive programs, actors must use \texttt{monitor}: whenever the monitored stream produces a new result, a new message is enqueued in the mailbox of the actor.
	Reactors have no imperative operators to ``send'' or ``receive'' values, or to manually react to the changes of values (e.g.~via callbacks).
	In the true spirit of reactive programming, they can only declaratively express dependencies on data sources via their source nodes and qualifications.
\end{description}

\subsection{Additional Mentions}
\label{subsec:evaluation-additional-mentions}
There are some reactive languages and frameworks that require a special mention.
When possible we list them in \cref{table:evaluation} below the double horizontal line.

Elm~\cite{DBLP:conf/pldi/CzaplickiC13} is a reactive programming language that compiles to JavaScript.
We were unable to build our running example in Elm because its current distribution is no longer reactive~\cite{elm-no-frp}. 
Expressions inside the DAG in Elm may perform infinite computations that block the reactive program.
An interesting observation is that Elm is presented as a purely functional reactive programming language, but its paper describes a \texttt{syncGet} operation to execute a web request (e.g.~to fetch an image from a URL), which is clearly a side-effect.
The reason why this operation (among others) is built-in is because it is \emph{necessary} for building web applications, but introducing this operation in the reactive language also causes the problems of the awkward squad.
This is exactly the essence of the awkward squad. 

ActiveSheets~\cite{DBLP:conf/ecoop/VaziriTRSH14} is a reactive language based on Microsoft Excel where the DAG consists of regular spreadsheet operations.
ActiveSheets adds features to Excel to automatically insert values into cells based on external data streams, and, like a regular spreadsheet, updates to cells automatically propagate throughout the program.
The core language of ActiveSheets is formalised and used to prove that, for any given update of a cell, computation time and memory usage are bound.
Furthermore, as far as we know there are no spreadsheet operations that have side-effects on other cells, so there can be no side-effects in internal nodes of the DAG.
However, ActiveSheets is not a general purpose programming language.
Conceptually an ActiveSheets program can be represented by 1 reactor that implements the spreadsheet logic.

In the Coherence language, code is divided in \textit{derivations} and \textit{reactions}~\cite{DBLP:conf/oopsla/Edwards09}.
Derivation is used to automatically compute the program output by deriving values from input via purely functional computations.
Side-effects are isolated to different parts of the code, namely reactions, that are responsible for imperatively keeping the application state consistent with derived values.
This stems from the insight that derivation and reaction need each other, but that coordinating side-effects in an event-based application is extremely difficult. 
A reactive program constructed via derivations is guaranteed to be free of side-effects, and is thus not subject to the Reactive Update Order Leak.
The Coherent Reaction programming model offers no mechanisms to solve the Reactive Thread Hijacking Problem.

HipHop is a synchronous reactive programming language inspired by Esterel with an implementation in Scheme~\cite{DBLP:conf/icdcit/BerryS14} and JavaScript~\cite{DBLP:conf/sac/VidalBS18}.
HipHop is not classified in \cref{table:evaluation} because the programming style and evaluation model of synchronous reactive programming languages makes them difficult to compare with the approaches in \cref{table:evaluation}.
However, there are interesting parallels between some aspects of the HipHop language and the Actor-Reactor Model.
HipHop is embedded as a DSL within the Hop language, and Hop code interfaces with HipHop code via \emph{reactive machines} that conceptually fulfil the same role as reactors. 
Hop code sends input events to a reactive machine and manually triggers a propagation cycle.
Output events produced by the HipHop machine can be observed by Hop code via event handlers.
Interestingly, one of the core language statements called \texttt{atom\&} is used within HipHop to execute Hop code, which may contain side-effects and recursive functions.
While side-effects in synchronous reactive programs are arguably not subject to the issues discussed in \cref{subsec:effectful-computations}, the authors make a note that the ``\textit{execution time [of atom statements] should be kept negligible in practice}''~\cite[3.4]{DBLP:conf/icdcit/BerryS14}.

\section{Conclusion}
To conclude this paper we reflect on our 2 largest contributions, namely identifying the awkward squad for reactive programming and the Actor-Reactor Model.
We believe that there is no panacea to write both imperative and reactive programs within a single unified language that exposes the same concepts and operations in both types of programs.
Instead, we believe that imperative and reactive programs are fundamentally different, and that they should be programmed whilst guaranteeing their own invariants.
To this end the Actor-Reactor Model serves as a new mental model to classify and design reactive systems.

As a secondary contribution, our definition of reactors may prove to be valuable to the field of reactive programming in two ways. 
First, we define distinct terminology for the different stages of a reactive program: a reactor behaviour represents a dependency graph that is constructed from code, a deployment is a specific instance of a graph that keeps track of all run-time information and state, and a reactor contains the reactive engine that propagates values through one or more deployments.
Conceptually, the reactive programs in many existing reactive programming languages are analogous to 1 reactor with 1 deployment.
By using our definitions, we open the door for modularity and composition of reactive programs both statically (e.g.~via point-wise and point-free composition operators) and dynamically by composing data streams.
Second, we conjecture that reactors together with the mechanism of qualification is an alternative, but equally powerful, way to construct higher-order reactive programs.
Reactors may provide more insights or clarities with respect to the run-time semantics and resource usage of higher order reactive programs.

The Actor-Reactor Model may inspire designers of reactive languages, framework developers, and researchers to be strict about what can and cannot be programmed within a reactive program, and it may help them to define precisely the rules by which reactive programs interact with their environment.
Additionally, the Actor-Reactor Model may be especially useful in application domains where certain parts of a program must be reactive, for example, with specific memory or timing constraints (e.g.~robotics and safety systems).
Operations that might not fit this model, but which are necessary for program development, are evacuated into actors which are complementary to the reactive program, but which do not violate the invariants of the reactive programming model.



\bibliography{references}

\end{document}